\documentclass{jetpl}
\twocolumn
\usepackage{amsfonts}
\usepackage{bm}
\usepackage{verbatim}
\usepackage{graphicx}

\lat

\title{Graphite, graphene  and the flat band superconductivity}

\rtitle{Graphite, graphene  and their superconductivity}

\sodtitle{Graphite, graphene  and their superconductivity}

\author{
G.E. Volovik\/\thanks{e-mail: volovik(at)ltl.ttk.fi}}

\sodauthor{Volovik}

\address{Low Temperature Laboratory, Aalto University,  P.O. Box 15100, FI-00076 Aalto, Finland\\~\\
Landau Institute for Theoretical Physics, acad. Semyonov av., 1a, 142432,
Chernogolovka, Russia
}

\dates{\today}{*}

\begin{document}

\maketitle

Superconductivity with transition temperature $T_c=1.7$ K has been observed  in bilayer 
graphene\cite{Cao2018a,Cao2018b}.  The main factors, which may shed light on the mechanism of the formation of this superconductivity, are the following.   Superconductivity is observed in bilayer graphene, when the two layers are twisted,  and the maximum of $T_c$ takes place at the "magic angle" of twist, at which the electronic band structure becomes nearly flat. 

Actually the same factors have been suggested in Ref. \cite{Esquinazi2014} to explain the experiments in graphite, which reported  high-T superconductivity in  highly oriented pyrolytic graphite (HOPG)  
\cite{Esquinazi2013a,Esquinazi2013b,Esquinazi2013c,Precker2016,Esquinazi2018}.
The hints of room-temperature superconductivity are present, only when the sample contains 
 quasi two-dimensional interfaces between the domains of HOPG.\cite{EsquinaziReply}
These domains should be twisted with respect to each other in order to form 
the flat band in electronic spectrum. \cite{Esquinazi2014} 
This   dispersionless energy spectrum has a singular density of states, which provides the transition
temperature being proportional to the coupling constant instead of the exponential 
suppression, as was suggested by Khodel and Shaginyan.\cite{KhodelShaginyan1990}
 For nuclear systems the linear dependence of the gap on the coupling constant has been
found by Belyaev \cite{Belyaev1961}.

There are different potential sources of the formation of the electronic flat band: due to topology
\cite{HeikkilaVolovik2016,HeikkilaKopninVolovik2011,Hyart2018,Pal2018}, due to symmetry \cite{Lieb1989,Bistritzer2011} and due to interactions \cite{KhodelShaginyan1990,Volovik1991,Nozieres1992,Yudin2014,Volovik1994,Dolgopolov2014}. 
The twist of the graphene layers or HOPG domains provides the topological mechanism of the flat band formation. The situation has the close relation to superconductivity at the interfaces or surfaces of topological crystalline semiconductors, see e.g.  \cite{TangFu2014}.
The reason of the formation of the nearly flat band can be the spontaneous formation of a misfit dislocation array at the interface. 
The topological origin of this flat band can be also understood in terms of the
pseudo-magnetic field created by strain \cite{Vozmediano2013,Heikkila2016,Ramires2018}.
The topologically protected band touching lines in the spectrum of graphite \cite{Mikitik2014,Mikitik2006,Mikitik2008}  (the so-called Dirac lines)
lead to approximate flat band on the surface of graphite or at the interface \cite{HeikkilaVolovik2016,HeikkilaKopninVolovik2011,Hyart2018}.
The Khodel-Shaginyan mechanism of flattening -- the merging of levels due to interaction  
\cite{KhodelShaginyan1990,Volovik1991,Nozieres1992,Yudin2014,Volovik1994,Dolgopolov2014,Dolgopolov2017} -- is also not excluded, since it leads to the further flattening of the 
spectrum.

Formation of the superconducting and ferromagnetic states in the flat band materials has been discussed in particular in Refs. \cite{Heikkila2016,Kopnin2011,Kopnin2013,Torma2017,Heikkila2018,Tasaki1992,Mielke1993,Lotman2017}.

At the moment there are many evidences of enhanced superconducting transition temperature in graphite materials, starting from Refs.\cite{Kopelevich1999,Kopelevich2000}.
 The most clear hints of high-T superconductivity in graphite are related to 
interfaces in HOPG
\cite{Esquinazi2014,Esquinazi2013a,Esquinazi2013b,Esquinazi2013c,Precker2016,Esquinazi2018}.
This supports the idea that  it is the twisted interface, which is the reason for the flattening of the spectrum. The electron-electron interaction
\cite{KhodelShaginyan1990,Volovik1991,Nozieres1992} is probably a secondary factor enhancing the flat band singularity in the electronic density of states.

Signatures of  high-T superconductivity are also observed in graphite in contact to other materials. Examples are alkanes in contact with a graphite surface \cite{Kawashima2013,Kawashima2018},
 polymer composites with embedded graphene flakes \cite{Ionov2015,Ionov2016}, 
 films of polystyrene and graphene oxide composite \cite{Khairullin2016},  sulfur doped amorphous carbon \cite{Kopelevich2001,Felner2016}, phosphorus-doped graphite and graphene \cite{Larkins2016},
 etc. Signatures of superconductivity with $T_c$=14 K have been found on the surface of Grafoil.\cite{Saunders2018}

In conclusion, the observation of superconductivity in graphene supports  the specific role played by twist of the graphite planes and by the band flattening. 
Graphite superconductivity is now becoming the mainstream. One may say that we are coming to graphite era of superconductivity. It is time to combine the theoretical and experimental efforts  to reach the bulk room-T superconductivity in graphite and in similar real or artificial materials.

 I thank T. Heikkil\"a for fruitful dicussions. This work has been supported by the European Research Council (ERC) under the European Union's Horizon 2020 research and innovation programme (Grant Agreement No. 694248).

\end{document}